# Foldy-Wouthuysen Transformation with Dirac Matrices in Chiral Representation

V.P.Neznamov

RFNC-VNIIEF, 607190, Sarov, Nizhniy Novgorod region

Abstract

The paper offers an expression of the general Foldy-Wouthuysen transformation in the chiral representation of Dirac matrices and in the presence of boson fields $B^\mu(\vec{x}, t)$ interacting with fermion field $\psi(\vec{x}, t)$.

The papers [1], [2] discuss the theory of interacting quantum fields in the Foldy-Wouthuysen representation [3]. These papers offer, in particular, the relativistic nonlocal Hamiltonian $H_{FW}$ in the form of a series in terms of powers of charge $e$. Quantum electrodynamics in the Foldy-Wouthuysen ($FW$) representation has been formulated using Hamiltonian $H_{FW}$ and some quantum electrodynamics processes have been calculated within the lowest-order perturbation theory. As a result, the conclusion has been made that the $FW$ representation describes some quasi-classic states in the quantum field theories. Both particles and antiparticles are available in these states. Particles, as well as antiparticles, interact with each other. However, there is no interaction of real particles with antiparticles – such interaction is possible only in intermediate (virtual) states. The $FW$ representation modification is required to take into account real particle/antiparticle interactions. In the papers [1], [2] such modification has been made using the symmetry identical to the isotopic spin symmetry owing to invariance of final physical results under change of signs in the mass terms of Dirac Hamiltonian $H_D$ and Hamiltonian $H_{FW}$. In the modified Foldy-Wouthuysen representation, real fermions and antifermions can be in two states characterized by the values of the third component of the isotopic spin $T_f^3 = \pm\frac{1}{2}$; real fermions and antifermions interacting with each other must have opposite signs of $T_f^3$. Quantum electrodynamics in the modified $FW$ representation is invariant under $P$-, $C$-, $T$-transformations. Violations of the introduced symmetry of the isotopic spin lead to the corresponding violation of $CP$-invariance. The Standard Model in the modified $FW$ representation was formulated in the papers [1], [4]. It has been shown that formulation of the theory in the modified $FW$ representation doesn't require that Higgs bosons should obligatorily interact with fermions to preserve the $SU(2)$-invariance, whereas all the rest theoretical and experimental implications of the Standard model obtained in the Dirac representation are preserved. In such a case, Higgs bosons are responsible only for the gauge invariance of the boson sector of the theory and interact only with gauge bosons $W_\mu^\pm, Z_\mu$, gluons and photons.

In the papers mentioned above, the energy representation of Dirac matrices derived by Dirac himself is used:

$$\alpha^i = \begin{pmatrix} 0 & \sigma^i \\ \sigma^i & 0 \end{pmatrix}, \beta = \gamma^0 = \begin{pmatrix} I & 0 \\ 0 & -I \end{pmatrix}, \gamma_5 = \begin{pmatrix} 0 & I \\ I & 0 \end{pmatrix}, \gamma^i = \gamma^0 \alpha^i. \qquad (1)$$

The question arises: what changes of the Foldy-Wouthuysen transformation form will result from using the chiral representation of Dirac matrices?

$$\alpha^i = \begin{pmatrix} \sigma^i & 0 \\ 0 & -\sigma^i \end{pmatrix}, \beta = \gamma^0 = \begin{pmatrix} 0 & I \\ I & 0 \end{pmatrix}, \gamma_5 = \begin{pmatrix} I & 0 \\ 0 & -I \end{pmatrix}, \gamma^i = \gamma^0 \alpha^i \qquad (2)$$

The chiral representation (2) is commonly used in the modern gauge field theories and in the Standard Model, in particular.

To answer the question above, first consider the structure of equations describing the components of the wave functions $\psi_D(x)$ for the two representations of Dirac matrices considered in the paper.

In relations (1), (2) and below the system of units with $\hbar = c = 1$ is used; $x$, $p$, $A$ are 4-vectors; the inner product is taken as $x^\mu y_\mu = x^0 y^0 - x^k y^k$, $\mu=0,1,2,3$, $k=1,2,3$; $p^\mu = i\dfrac{\partial}{\partial x_\mu}$; $\sigma^k$ are Pauli matrices; $\alpha^\mu = \begin{cases} 1, \mu = 0 \\ \alpha^i, \mu = k, k=1,2,3 \end{cases}$; $\psi_D(x)$ is the four-component wave function, $\varphi(x), \chi(x), \psi_R(x), \psi_L(x)$ are the two-component wave functions.

The following operator relations are valid for the free Dirac equation with representation (1):

$$p_0 \psi_D(x) = (\vec{\alpha}\vec{p} + \beta m)\psi_D(x); \psi_D(x) = \begin{pmatrix} \varphi(x) \\ \chi(x) \end{pmatrix};$$

$$\begin{cases} p_0 \varphi(x) = \vec{\sigma}\vec{p}\chi(x) + m\varphi(x) \\ p_0 \chi(x) = \vec{\sigma}\vec{p}\varphi(x) - m\chi(x) \end{cases}; \qquad (3)$$

$$\chi = (p_0 + m)^{-1} \vec{\sigma}\vec{p}\varphi;$$
$$\varphi = (p_0 - m)^{-1} \vec{\sigma}\vec{p}\chi$$

With representation (2), relation (3) looks like

$$p_0 \psi_D(x) = (\vec{\alpha}\vec{p} + \beta m)\psi_D(x); \psi_D(x) = \begin{pmatrix} \psi_R(x) \\ \psi_L(x) \end{pmatrix};$$

$$\begin{cases} p_0 \psi_R(x) = \vec{\sigma}\vec{p}\,\psi_R(x) + m\psi_L(x) \\ p_0 \psi_L(x) = -\vec{\sigma}\vec{p}\,\psi_L(x) + m\psi_R(x) \end{cases};  \tag{4}$$

$$\psi_L(x) = \frac{p_0 - \vec{\sigma}\vec{p}}{m}\psi_R(x) = (p_0 + \vec{\sigma}\vec{p})^{-1} m\psi_R(x)$$

$$\psi_R(x) = \frac{p_0 + \vec{\sigma}\vec{p}}{m}\psi_L(x) = (p_0 - \vec{\sigma}\vec{p})^{-1} m\psi_L(x)$$

Relations (4) use the operator equality:

$$p_0^2 = E^2 = \vec{p}^{\,2} + m^2 .$$

Comparison between relations (3) and (4) shows that with the substitution below,

$$m \leftrightarrow \vec{\sigma}\vec{p}, \ \beta \leftrightarrow \gamma_5  \tag{5}$$

these relations transform into each other.

The Foldy-Wouthuysen transformations for the energy and chiral representations of Dirac matrices also transform into each other, if substitution (5) is made.

The Foldy-Wouthuysen transformations for free motion of fermions (without boson fields $B^\mu(x)$) look like

$$\left(U_{FW}^0\right)^{en} = \sqrt{\frac{E+m}{2E}}\left(1 + \frac{\beta\gamma_5\vec{\sigma}\vec{p}}{E+m}\right)  \tag{6}$$

$$\left(U_{FW}^0\right)^{chir} = \sqrt{\frac{E+\vec{\sigma}\vec{p}}{2E}}\left(1 + \frac{\gamma_5\beta m}{E+\vec{\sigma}\vec{p}}\right)  \tag{7}$$

Expression (6) is written for the energy representation of Dirac matrices; expression (7) is the desired one describing the Foldy-Wouthuysen transformation with Dirac matrices in the chiral representation; in expressions (6), (7) $E = \sqrt{\vec{p}^{\,2} + m^2}$.

Matrices (6), (7) are unitary and

$$\left(U_{FW}^0\right)^{en}(\vec{\alpha}\vec{p} + \beta m)\left(U_{FW}^0\right)^{en\dagger} = \beta E;  \tag{8}$$

$$\left(U_{FW}^0\right)^{chir}(\vec{\alpha}\vec{p} + \beta m)\left(U_{FW}^0\right)^{chir\dagger} = \gamma_5 E;  \tag{9}$$

If boson fields $B^\mu(x)$ interacting with fermion field $\psi(x)$ are present, the desired Foldy-Wouthuysen transformation with Dirac matrices in the chiral representation ($U_{FW}^{chir}$) and the fermion Hamiltonian ($H_{FW}^{chir} = \gamma_5 E + qK_1 + q^2K_2 + q^3K_3 + ...$; $q$ is the coupling constant) of the corresponding form can be obtained using the algorithm described in the Refs.[1], [2] along with substitution (5).

For example, expressions for operators $C$ and $N$ forming the basis for Hamiltonian of interaction in the Foldy-Wouthuysen representation (see [1], [2]) can be written in the following form for Dirac matrices in the chiral representation:

$$C = \left[ (U_{FW}^0)^{chir} q\alpha_\mu B^\mu (U_{FW}^0)^{chir\dagger} \right]^{even} = qR(B^0 - LB^0L)R - qR(\vec{\alpha}\vec{B} - L\vec{\alpha}\vec{B}L)R;$$

$$L = \frac{\gamma_5 \beta m}{E + \vec{\sigma}\vec{p}}, \quad R = \sqrt{\frac{E + \vec{\sigma}\vec{p}}{2E}} \quad (10)$$

$$N = \left[ (U_{FW}^0)^{chir} q\alpha_\mu B^\mu (U_{FW}^0)^{chir\dagger} \right]^{odd} = qR(LB^0 - B^0L)R - qR(L\vec{\alpha}\vec{B} - \vec{\alpha}\vec{B}L)R \quad (11)$$

*even, odd* in expressions (10), (11) denote the even and odd parts of the corresponding operator (see [1], [2]).

Thus, the general Foldy-Wouthuysen transformation with Dirac matrices in the chiral representation $U_{FW}^{chir} = (U_{FW}^o)^{chir}(1 + \delta_1^{chir} + \delta_2^{chir} + \delta_3^{chir} + ...)$, as well as the fermion Hamiltonian in the Foldy-Wouthuysen representation

$$H_{FW}^{chir} = \gamma_5 E + qK_1^{chir} + q^2 K_2^{chir} + q^3 K_3^{chir} + ...$$

can be obtained from the corresponding expressions for $U_{FW}^{en}, H_{FW}^{en}$ with Dirac matrices in the energy representation (see [1], [2]) with substitution $m \leftrightarrow \vec{\sigma}\vec{p}, \beta \leftrightarrow \gamma_5$.

Certainly, physical results do not depend on the chosen representation of Dirac matrices. A researcher chooses a certain representation for the sake of convenience.

The Appendix below gives examples of calculations for two quantum eletrodynamics processes in Foldy-Wouthusyen representation using Dirac matrices in the chiral representation.

**Appendix**

# Calculation of some quantum electrodynamics processes in FW representation with Dirac matrices in the chiral representation

The matrix elements and calculations below are given using the notations from [1], [2].

1. Electron scattering in Coulomb field is $A_0(x) = -\dfrac{Ze}{4\pi|\vec{x}|}$.

Feynman diagram of the process is shown in Fig.1.

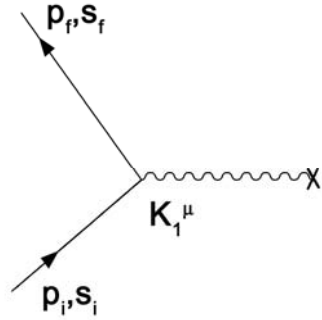

Figure 1. Electron scattering in Coulomb field

$$S_{fi} = -i\int d^4x (\Psi_{FW}^{(+)}(x,p_f,s_f))^\dagger K_1^0 \cdot A_0 \Psi_{FW}^{(+)}(x,p_i,s_i) =$$

$$= -\frac{i\delta(E_f - E_i)}{(2\pi)^2} U_{s_f}^\dagger <\vec{p}_f | C^o A_0 | \vec{p}_i > U_{s_i} =$$

$$= i\frac{Ze^2}{\vec{q}^{\,2}} \frac{\delta(E_f - E_i)}{(2\pi)^2} U_{s_f}^\dagger \sqrt{\frac{E_f + \vec{\sigma}\vec{p}_f}{2E_f}} \left(1 - \frac{\gamma^5 \beta m}{E_f + \vec{\sigma}\vec{p}_f} \frac{\gamma^5 \beta m}{E_i + \vec{\sigma}\vec{p}_i}\right) \sqrt{\frac{E_i + \vec{\sigma}\vec{p}_i}{2E_i}} U_{s_i}, \vec{q} = \vec{p}_f - \vec{p}_i.$$

The entry $K_1^0 \cdot A_0$ made for the sake of convenience actually means that $K_1^0 \cdot A_0 \equiv K_1$ with $\vec{A}(x) = 0$. In other words, $A_0(x)$ occupies the positions determined by expression (26) from the paper [1]. The same is valid for the entry $C^0 \cdot A_0$. The transition from $K_1^0 \cdot A_0$ to $C^0 \cdot A_0$ has been made according to expression (39) from the paper [1].

Then, having the matrix element $S_{fi}$ and using the common methods one can obtain the differential Mott scattering cross-section that takes the form of the Rutherford cross-section in non-relativistic case.

$$\frac{d\sigma}{d\Omega} = \frac{4Z^2\alpha^2 E^2}{2\vec{q}^4} Sp\left\{ \sqrt{\frac{E+\vec{\sigma}\vec{p}_f}{2E}} \left(1 + \frac{m^2}{(E+\vec{\sigma}\vec{p}_f)(E+\vec{\sigma}\vec{p}_i)}\right) \frac{E+\vec{\sigma}\vec{p}_i}{2E} \left(1 + \frac{m^2}{(E+\vec{\sigma}\vec{p}_i)(E+\vec{\sigma}\vec{p}_f)}\right) \right.$$
$$\left. \times \sqrt{\frac{E+\vec{\sigma}\vec{p}_f}{2E}} \right\};$$

$$E = E_i = E_f, p = |\vec{p}_f| = |\vec{p}_i|, \vec{p}_f \vec{p}_i = p^2 \cdot \cos\theta, \beta = \frac{p}{E}, \alpha = \frac{e^2}{4\pi};$$

Further transformations give us the differential Mott scattering cross-section:

$$\frac{d\sigma}{d\Omega} = \frac{Z^2\alpha^2}{4p^2\beta^2 \sin^4\frac{\theta}{2}} \left(1 - \beta^2 \sin^2\frac{\theta}{2}\right).$$

2. The self-energy of electron in the second-order perturbation theory. Feynman diagrams of the processes are shown in Fig. 2.

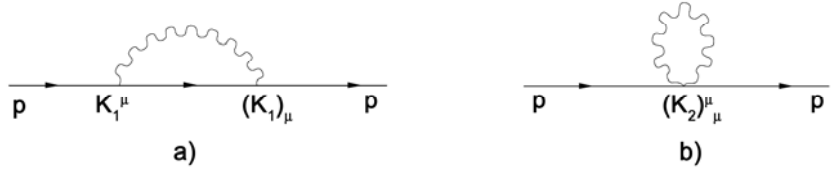

Figure 2. The electron self-energy

$$-i\sum\nolimits^{(2)}(p) = -\int \frac{d^4k}{(2\pi)^4 \cdot k^2}\Bigg[K_1^\mu\left(\vec{p};\vec{p}-\vec{k};v=-1\right)\frac{1}{p_0-k_0-\beta E\left(\vec{p}-\vec{k}\right)}K_{1\mu}\left(\vec{p}-\vec{k};\vec{p};v'=1\right)-$$

$$-K_1^\mu(\vec{p};\vec{p}-\vec{k};v=-1)\frac{\frac{1}{2}(1+\beta)}{\beta E(\vec{p})-k_0-E(\vec{p}-\vec{k})}\cdot K_{1\mu}(\vec{p}-\vec{k};\vec{p};v'=1)+$$

$$+C^\mu(\vec{p};\vec{p}-\vec{k})\frac{\frac{1}{2}(1+\beta)}{\beta E(\vec{p})-k_0-E(\vec{p}-\vec{k})}\cdot C_\mu(\vec{p}-\vec{k};\vec{p})+$$

$$+N^\mu(\vec{p};\vec{p}-\vec{k})\frac{\frac{1}{2}(1-\beta)}{-\beta E(\vec{p})-k_0+E(\vec{p}-\vec{k})}N_\mu(\vec{p}-\vec{k};\vec{p})\Bigg]$$

In view of $\beta\psi_{FW}^{(+)} = \psi_{FW}^{(+)}$,) for a free electron ($p^2=m^2$) the first two items in the integrand are mutually annihilating and expressions $C^\mu(\vec{p};\vec{p}_1)C_\mu(\vec{p}_1;\vec{p})$, $N^\mu(\vec{p};\vec{p}_1)N_\mu(\vec{p}_1;\vec{p})$ have the form

$$C^\mu(\vec{p};\vec{p}_1)C_\mu(\vec{p}_1;\vec{p}) =$$
$$= \sqrt{\frac{E+\vec{\sigma}\vec{p}}{2E}}\left(1+\frac{m^2}{(E+\vec{\sigma}\vec{p})(E_1+\vec{\sigma}\vec{p}_1)}\right)\left(\frac{E_1+\vec{\sigma}\vec{p}_1}{2E_1}\right)\left(1+\frac{m^2}{(E_1+\vec{\sigma}\vec{p}_1)(E+\vec{\sigma}\vec{p})}\right)\sqrt{\frac{E+\vec{\sigma}\vec{p}}{2E}} -$$
$$-\sqrt{\frac{E+\vec{\sigma}\vec{p}}{2E}}\left(\alpha^i - \frac{m}{(E+\vec{\sigma}\vec{p})}\alpha^i\frac{m}{(E_1+\vec{\sigma}\vec{p}_1)}\right)\left(\frac{E_1+\vec{\sigma}\vec{p}_1}{2E_1}\right)\left(\alpha^i - \frac{m}{(E_1+\vec{\sigma}\vec{p}_1)}\alpha^i\frac{m}{(E+\vec{\sigma}\vec{p})}\right)\times$$
$$\times\sqrt{\frac{E+\vec{\sigma}\vec{p}}{2E}} = \frac{1}{EE_1}\left(-EE_1 + \vec{p}\vec{p}_1 + 2m^2\right),$$

$$\vec{p}_1 = \vec{p}-\vec{k},\ E_1 = \sqrt{m^2+\left(\vec{p}-\vec{k}\right)^2};$$

$$N^\mu(\vec{p};\vec{p}_1)N_\mu(\vec{p}_1;\vec{p}) =$$
$$= \sqrt{\frac{E+\vec{\sigma}\vec{p}}{2E}}\left(\frac{\gamma_5\beta m}{(E+\vec{\sigma}\vec{p})}-\frac{\gamma_5\beta m}{(E_1+\vec{\sigma}\vec{p}_1)}\right)\left(\frac{E_1+\vec{\sigma}\vec{p}_1}{2E_1}\right)\left(\frac{\gamma_5\beta m}{(E_1+\vec{\sigma}\vec{p}_1)}-\frac{\gamma_5\beta m}{(E+\vec{\sigma}\vec{p})}\right)\sqrt{\frac{E+\vec{\sigma}\vec{p}}{2E}} -$$
$$-\sqrt{\frac{E+\vec{\sigma}\vec{p}}{2E}}\left(\frac{\gamma_5\beta m}{E+\vec{\sigma}\vec{p}}\alpha^i - \alpha^i\frac{\gamma_5\beta m}{E_1+\vec{\sigma}\vec{p}_1}\right)\left(\frac{E_1+\vec{\sigma}\vec{p}_1}{2E_1}\right)\left(\frac{\gamma_5\beta m}{E_1+\vec{\sigma}\vec{p}_1}\alpha^i - \alpha^i\frac{\gamma_5\beta m}{E+\vec{\sigma}\vec{p}}\right)\times$$
$$\times\sqrt{\frac{E+\vec{\sigma}\vec{p}}{2E}} = -\frac{1}{EE_1}\left(EE_1 + \vec{p}\vec{p}_1 + 2m^2\right).$$

Hence,

$$-i\sum\nolimits^{(2)}(p) = -\frac{2e^2}{E(\vec{p})}\int\frac{d^4k}{(2\pi)^4 k^2}\cdot\frac{pk+m^2}{[(p-k)^2-m^2]}$$

and, with regard to spinor normalization, this expression is similar to the expression describing the mass operator in Dirac representation.